\documentclass{article}
\usepackage{spconf,amsmath,graphicx,tikz,url}
\usetikzlibrary{spy}

\def\x{{\mathbf x}}
\def\s{{\mathbf s}}

\def\y{{\mathbf y}}
\def\n{{\mathbf n}}
\def\z{{\mathbf z}}

\def\d{{\mathbf d}}
\def\A{{\mathbf A}}

\title{A NOVEL METHOD AND DATASET FOR DEPTH-GUIDED IMAGE DEBLURRING FROM SMARTPHONE LIDAR}
\name{Antonio Montanaro, Diego Valsesia\thanks{This study was carried out within the ``AI-powered LIDAR fusion for next-generation smartphone cameras (LICAM)'' project – funded by European Union – Next Generation EU  within the PRIN 2022 program (D.D. 104 - 02/02/2022 Ministero dell’Università e della Ricerca), CUP E53D23000790006. This manuscript reflects only the authors’ views and opinions and the Ministry cannot be considered responsible for them.}}
\address{Politecnico di Torino, Italy}

\begin{document}
%
\maketitle
\begin{abstract}
Modern smartphones are equipped with Lidar sensors providing depth-sensing capabilities. Recent works have shown that this complementary sensor allows to improve various tasks in image processing, including deblurring. However, there is a current lack of datasets with realistic blurred images and paired mobile Lidar depth maps to further study the topic. At the same time, there is also a lack of blind zero-shot methods that can deblur a real image using the depth guidance without requiring extensive training sets of paired data. In this paper, we propose an image deblurring method based on denoising diffusion models that can leverage the Lidar depth guidance and does not require training data with paired Lidar depth maps. We also present the first dataset with real blurred images with corresponding Lidar depth maps and sharp ground truth images, acquired with an Apple iPhone 15 Pro, for the purpose of studying Lidar-guided deblurring. Experimental results on this novel dataset show that Lidar guidance is effective and the proposed method outperforms state-of-the-art deblurring methods in terms of perceptual quality.  
\end{abstract}
\section{Introduction}
\label{sec:intro}
Modern smartphones are increasingly becoming multimodal imaging devices with the inclusion of Lidar sensors in recent consumer devices, such as the Apple iPhone. Despite its limited resolution due to space and cost constraints, the active Lidar instrument can serve as complementary source of information to passive optical cameras, even in tasks that do not explicitly deal with 3D reconstruction. In fact, promising results have been shown in terms of enhanced rate-distortion performance for Lidar-guided image compression \cite{gnutti2024lidar} and improved image quality for Lidar-guided deblurring \cite{yi2024deep}. 

Focusing on deblurring, the active nature of the Lidar sensor can be useful in low-light scenarios where motion blur is easily introduced in photos, while depth information can provide sharp object boundaries.
While the work in \cite{yi2024deep} has shown that smartphone Lidar data can substantially enhance image deblurring, it has done so via simulation of blurred images, albeit with realistic kernels. In fact, currently, the main limitation towards a deeper development of the topic is the lack of curated data with real blurred images under low-light conditions, paired with Lidar depth maps. The only existing dataset with paired mobile Lidar depth maps is ARKitScenes \cite{dehghan2021arkitscenes}, which does not have real blurred images. Another limitation of \cite{yi2024deep} is the supervised training approach, which requires a large dataset with paired data, including blurred image, registered Lidar depth maps and sharp ground truth. It would be desirable to devise a zero-shot method that learns a general ``sharp image prior'' and side-information fusion operator that allows to avoid the dependency on large quantities of images with associated Lidar depth maps.  

In this paper, we address both of the aforementioned limitations. First, we present a novel dataset of real low-light images affected by motion blur, captured with an Apple iPhone 15 Pro smartphone, with registered Lidar depth maps and sharp ground truth images. Moreover, we present a novel blind zero-shot method for Lidar-guided image deblurring, called ZSLDB (Zero-Shot Lidar DeBlur) that leverages denoising diffusion generative models to avoid the need for extensive paired datasets required by supervised learning. The model is blind to the real degradation kernel, which is estimated during inference. It also allows for conditioning based on the Lidar depth maps without explicit training with paired data having them. Experimental results on the novel dataset of real blurred images with mobile Lidar depth maps show that the depth information improves the quality of the reconstructed images and that ZSLDB outperforms state-of-the-art algorithms in terms of perceptual image quality. 

Our novel contributions can be therefore summarized as:
\begin{itemize}
    \item we present a novel blind zero-shot method for image deblurring guided by Lidar depth maps, called ZSLDB;
    \item we introduce a novel dataset of real smartphone images affected by motion blur, with registered Lidar depth maps and sharp ground truth images;
    \item experimental results show that thanks to Lidar depth information ZSLDB achieves state-of-the-art perceptual quality;
    \item code and dataset are available at \url{https://github.com/diegovalsesia/ZSLDB}.
\end{itemize}

\section{Background}
\label{sec:bkg}

Image restoration, including deblurring, is a longstanding problem in the image processing field. Several solutions have been developed over the years, including optimization-based methods carefully modeling priors suitable to describe natural images. Deep learning has then shown that neural networks could effectively capture more sophisticated representations that led to improved reconstruction performance. At a high level, deep learning approaches generally either use a supervised learning strategy, directly learning the inverse mapping between degraded observations and reconstructions, or a generative model learning a data prior, including learned denoisers in plug-and-play approaches \cite{kamilov2023plug}, variational autoencoders (VAEs) \cite{kingma2013auto}, generative adversarial networks \cite{goodfellow2014generative}, denoising diffusion models \cite{song2019generative}, etc.

The former approach has proved to be very successful with recent models like Restormer \cite{zamir2022restormer}, NAFNet \cite{chen2022simple} achieving impressive results on a variety of inverse problems. Concerning blind image deblurring, where the blurring operator is not known in advance, the recent J-MKPD model \cite{carbajal2023blind} can be regarded as the state-of-the-art. 
The latter approach based on generative models is particularly suitable when image restoration has to face with real data, and it is difficult to source large quantities of paired data for supervised training. Morevoer, by modeling the real data distribution, methods based on generative models favour better perceptual quality in the perception-distrotion tradeoff. 
Examples for the solutions of inverse problems include works like PULSE \cite{menon2020pulse}, based on GAN inversion to seek a solution to the problem in the latent space of the generative model. DeblurGAN \cite{kupyn2018deblurgan} is another well-known work focusing on blind motion deblurring using a Wasserstein GAN. 

Since the rise of denoising diffusion models as the new state of the art in image generation, zero-shot methods exploiting these priors have become even more popular. In DPS \cite{chung2022diffusion}, the authors extend diffusion solvers to efficiently handle general noisy (non)linear inverse problems via approximation of the posterior sampling; DDRM \cite{kawar2022denoising}  reformulates the diffusion process to be guided by the degraded observation; in DDNM \cite{wang2022zero}, the authors decompose the restored signal in two algebraic parts and exploit the diffusion process to fill up the information of one part; \cite{zhu2023denoising} integrates a traditional plug-and-play method into the diffusion sampling framework; the authors in \cite{rout2024solving} are the first who consider latent diffusion model rather than pixel-space diffusion models, and they define a new correction term to DPS. Other works also utilize backpropagation along the sampling chain for image restoration \cite{song2023solving}. 
Some limitations of the previous works that challenge them on real-world data are the often-used assumption that the degradation operator is known (non-blind problem), or the use pixel-space diffusion models trained on simple datasets like Imagenet or faces FFHQ, which do not scale well to realistic natural images. 

Moreover, works in the literature typically do not assume the availability of side information to condition the reconstruction process, which is the case of our setting where the Lidar depth map should provide guidance about edges. Before the advent of deep learning, an effective solution towards side information guidance was the use of the guided filter \cite{he2012guided}. More recently, ControlNet \cite{zhang2023adding} has shown a general way to condition the neural networks used in denosing diffusion models with side information. To the best of our knowledge, there is no work about blind image restoration using a diffusion model guided by depth maps.


\section{Method}
\label{sec:method}
The overall goal of this paper is to investigate the problem of image deblurring from smartphone cameras, when side information in the form of a depth map acquired by the Lidar on the same device is available. Compared to \cite{yi2024deep} that worked on synthetically degraded data, we focus on real data by both devising a novel depth-guided zero-shot deblurring method based on diffusion models, which can deblur our target images without requiring training data with blurred images and Lidar depth maps, as well as a novel dataset presented in Sec. \ref{sec:dataset}.

\subsection{Problem setting, diffusion models and conditional control}

Image deblurring can be regarded as a linear inverse problem:
\begin{align}
\y = \A\x + \n
\end{align}
where the original sharp image $\x$ has been corrupted by a forward operator $\A$ and, possibly, an additive random noise $\n$ to generate the blurred image $\y$. In its simplest form, the forward operation is a convolution with a low-pass filter, representing a spatially-invariant degradation. In more complex models, a spatially-variant filter is used to capture nonstationary degradations.

The classical formulation to solve this ill-posed problem is optimization as a Maximum a Posteriori (MAP) estimation problem:
\begin{align*}
    \hat{\x} = \arg\min_{\x} \Vert \y - \A\x \Vert_2^2 + \lambda R(\x),
\end{align*}
A regularizer function $R(\x)$ captures the data prior by encoding the knowledge about the sharp original images.

With the advent of powerful generative models, strong image priors are available to regularize inverse problems.
In essence, many generative models learn a map $G$ from a simple distribution on a latent space $\mathcal{Z}$ to the real data distribution. This allows to solve an inverse problem via inversion to the latent space, i.e., seeking the point in the model latent space (thus guaranteeing a realistic generation) that best fits the degraded observations. More formally, the reconstructed image $\hat{\x}$ is obtained as:
\begin{align} 
\hat{\x} &= G(\hat{\z}) \\
\hat{\z} &= \arg\min_{\z} \Vert \y - \A G(\z) \Vert_{2}^{2} \quad \text{s.t.} \quad z \in \mathcal{Z} .
\label{eq:inversion}\end{align}


Denoising Diffusion Probabilistic Models (DDPMs) \cite{song2019generative} are generative models that utilize a noise diffusion process to model the image distribution starting from random noise of the same dimension. The model is trained to progressively denoise data with different levels of noise, allowing the model to generate new data from pure noise by reverting the forward process; this is denoted as backward process and can be formulated as:
\begin{multline}
    p_\theta(\x_{0:T}) = p(\x_T) \prod\nolimits_{t=1}^T p_\theta(\x_{t-1} \mid \x_t) \qquad \\
    p_\theta(\x_{t-1} \mid \x_{t}) =\mathcal{N}(\x_{t-1} \mid \mu_{\theta}({\x}_{t}, t), \sigma_t^2\mathbf{I}). 
\end{multline}
where $\mu_{\theta}({\x}_{t}, t)$ is a function of a denoising neural network parametrized by $\theta$. A widely used neural network architecture for the denoiser is a UNet with self-attention blocks \cite{ronneberger2015u}.
Even if this process is Markovian, the work of \cite{song2020denoising} shows that is possible to construct a non-Markovian process defining a faster deterministic sampler (DDIM) that is compatible with a pretrained model.
So starting from $p_\theta(\x_{0:T})$, it is possible to sample $\x_{t-1}$ using:
\begin{multline}
    x_{t-1} = \sqrt{\alpha_{t-1}} \left(\frac{x_t - \sqrt{1-\alpha_t}\hat{\epsilon}_t}{\sqrt{\alpha_t}}\right) \\ + \sqrt{1-\alpha_{t-1}-\sigma_{t}(\eta)^2} \cdot \hat{\epsilon}_t+\sigma_{t}(\eta)\varepsilon_t
\end{multline}
where $\sigma_{t}(\eta) = \eta \sqrt{\frac{1-\alpha_{t-1}}{1-\alpha_{t}}}\sqrt{\frac{1-\alpha_{t}}{\alpha_{t-1}}}$, $\varepsilon_t$ is a normalized gaussian variable and $\epsilon_t(\cdot)$ is the denoiser neural network estimating the noise realization.
$\eta \in (0,1)$ is a parameter controlling the forward process. When $\eta = 0$ the sampling becomes deterministic, enabling the inversion approach described in Eq. \eqref{eq:inversion}.
Stable Diffusion is a latent diffusion model \cite{rombach2022high} that uses a Variational Auto-Encoder (VQ-VAE \cite{van2017neural}) to reduce the image dimensions and run the diffusion process in the reduced space encoded by the VAE encoder $E$, and after the final denoising step, let the VAE decoder $D$ recover the full image.

Diffusion models can be easily extended to model $p(\x | \s)$, where $\s$ is a conditioning signal, such as an
image caption, category, semantic maps etc., either by providing an additional input to the denoising neural network, which is typically used for text-conditioned generation, or by means of subnetworks that modulate the features of the denoiser. The latter approach has been shown by ControlNet \cite{zhang2023adding} to allow the integration of various forms of side information without the need to retrain the denoiser of the generative model. 

In the next section, we show how we use these concepts to design a zero-shot Lidar-guided deblurring method. 

\begin{figure*}[t]
    \centering
    \includegraphics[width=0.33\linewidth]{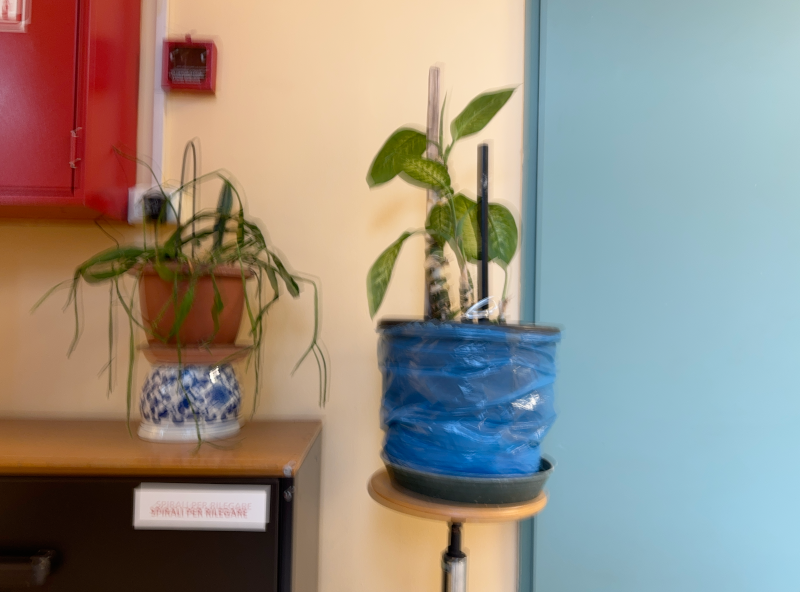}
    \includegraphics[width=0.33\linewidth]{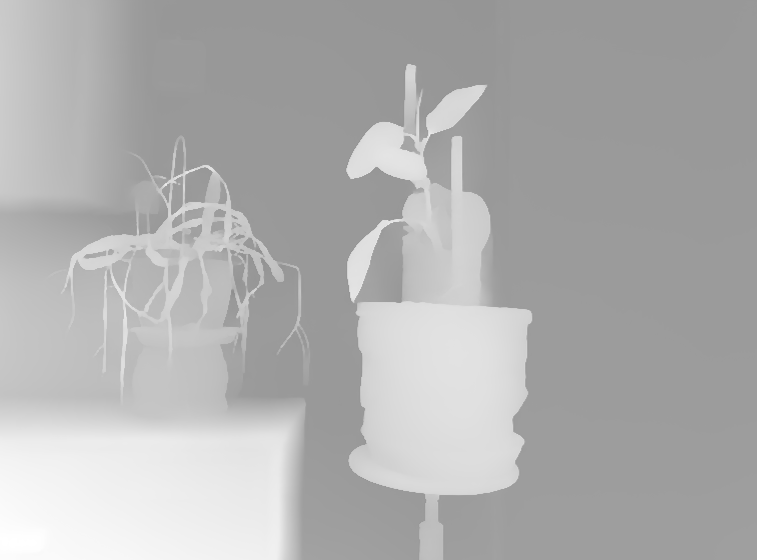}
    \includegraphics[width=0.33\linewidth]{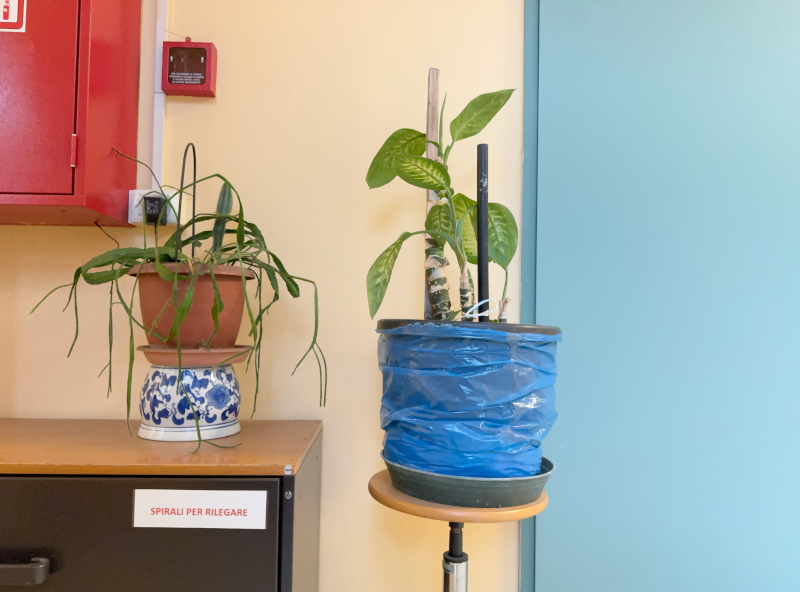}
    \vspace{-20pt}
    \caption{Sample scene from the novel smartphone Lidar-guided deblurring dataset. Left to right: blurred scene, Lidar depth map, sharp ground truth. Acquired with an Apple iPhone 15 Pro.}
    \label{fig:data_example}
\end{figure*}

\subsection{Zero-Shot Lidar DeBlur (ZSLDB)}
We propose a method called Zero-Shot Lidar DeBlur (ZSLDB) to leverage a pretrained diffusion model to deblur real images acquired by smartphones while conditioning the results on the information of the Lidar depth map captured by the same device. We frame the problem as a blind deblurring problem, as the blur kernel $\A$ is unknown in real images. Our experiments are based on using StableDiffusion (SD) as generative model, but the formulation extends to any latent diffusion model. We denote the Lidar depth map to be used as side information as $\d$ and the mapping performed by the diffusion model through the iterated denoiser function as $G$. 

Conceptually, ZSLDB casts the deblurring problem as the solution to the following optimization problem, following the inversion principle outlined in Eq. \eqref{eq:inversion}:

\begin{align}\label{eq:min_z_A}
\hat{\z}, \hat{\A} = \arg\min_{\z, \A} \Vert \y - \A G(\z,\d) \Vert_{2}^{2},\quad \hat{\x} = G(\hat{\z},\d).
\end{align}

Once the optimization finds the desired latent noise $\hat{\z}$, this is used by the model to generate the deblurred image $\hat{\x}$.
Notice that the optimization is both over the noise latent $\z$ of the SD model and the blur kernel $\A$. The diffusion model $G$ is a function of the depth signal by means of a ControlNet, mapping a resampled version of the depth map to match the image resolution, to residual features that are added to the denoiser features.

The optimization in Eq. \eqref{eq:min_z_A} is performed as follows. We start from the blurred image $\y$ and we use DDIM (50 timesteps) to invert this image back to the SD noise latent space at $T=0$. This inversion has no information about the depth map yet. Then, we generate an image starting from this latent by adding $\d$ as input to the ControlNet and we run the DDIM sampler with only 10 timesteps to generate a guided version of $\y$. Finally, we optimize the latent $\z$ by backpropagation through the sampling chain, adding some regularizers to basic idea of Eq. \ref{eq:min_z_A} and by measuring consistency with the observations in the latent space of the VAE model:
\begin{multline}\label{eq:stageI_loss}
\hat{\z}, \hat{\A} = \arg\min_{\z, \A} \Big[ \Vert E(\y) - E(\A G(\z,\d)) \Vert_{2}^{2} \\+ \gamma L_\text{aesthetic}(G(\z,\d)) + \lambda \cdot \text{LPIPS}(\y,G(\z,\d)) \Big].
\end{multline}
We found that optimization is more stable and effective by measuring fidelity with the blurred observations both as L2 norm in the VAE latent space, i.e., after the image encoder $E$, and with the LPIPS \cite{zhang2018unreasonable} perceptual loss in the pixel space. Additionally, $L_\text{aesthetic}$ is an aesthetic differentiable reward function as formulated in \cite{prabhudesai2023aligning}. This metric was devised by training a multi-layer perceptron operating on CLIP embeddings of 176,000 images to regress human ratings of perceptual quality. 
The hyperparameters $\gamma$ and $\lambda$ control the strength of the two regularizations and are set to 0.1 and 1.5, respectively.
Finally, we remark that backpropagation through the sampling chain of the diffusion process is made computationally feasible by means of the commonly-used gradient checkpointing \cite{prabhudesai2023aligning,gruslys2016memory} technique to lower memory requirements. We remark that the first inversion process uses a higher number of timesteps to ensure an accurate estimation of the latent noise, while subsequent DDIM sampling uses fewer timesteps for the purpose of acceleration, without significantly reducing image quality.

\section{Dataset}
\label{sec:dataset}

Current research efforts into the investigation of the use of smartphone Lidar sensors for the regularization of inverse problems in imaging are limited by the lack of available data. In particular, the only existing dataset with paired smartphone images and Lidar depth maps captured by the same device is ARKitScenes \cite{dehghan2021arkitscenes}. However, this dataset was created for 3D reconstruction and understanding tasks and not for image restoration. As such, it does not have images affected by realistic degradations such as motion blur.

In this paper, we present a novel dataset acquired with an Apple iPhone 15 Pro of low-light images affected by motion blur. We use the onboard Lidar sensor to also obtain registered depth maps. A sharp ground truth acquired by the same device in a more stable manner is also available and registered to the blurred image and depth map. The dataset is composed of 45 scenes, mostly indoors. We argue that indoor usage is the scenario where the Lidar sensor can be most effective since objects with detectable boundaries will be in range of the instrument. The dataset also comprises acquisitions of the same scenes from a DSLR camera. However, these are only coarsely registered due to differences in focal length but can be used for distribution-level quality assessment rather than paired metrics.

An example of a blurred scene, Lidar depth map and sharp ground truth is shown in Fig. \ref{fig:data_example}. The full dataset will be made publicly available upon publication.

\section{Experimental Results}
\label{sec:exp}

\begin{table}[t]
\centering
\caption{Quantitative deblurring results}
\vspace{6pt}
\label{table:main_exp}
\begin{tabular}{lc}
\textbf{Method} & \textbf{LPIPS $\downarrow$} \\
\hline
DeblurGAN \cite{kupyn2018deblurgan} & 0.1884 \\
Restormer \cite{zamir2022restormer} & 0.1783 \\
J-MKPD \cite{carbajal2023blind} & 0.1819 \\
\textbf{ZSLDB (ours)} & \textbf{0.1643}  \\
\hline
\end{tabular}
\end{table}

\begin{table}[t]
\centering
\caption{Impact of depth information}
\vspace{2pt}
\label{table:depth}
\begin{tabular}{ccc}
      & ZSLDB no depth  & \textbf{ZSLDB} \\
      \hline
LPIPS $\downarrow$ &    0.1821    & \textbf{0.1643}  \\
\hline
\end{tabular}
\end{table}

\begin{table}[t]
\centering
\caption{Conditioning on depth v. edge map}
\vspace{2pt}
\label{table:canny}
\begin{tabular}{ccc}
      & ZSLDB (depth)  & ZSLDB (edge) \\
      \hline
LPIPS $\downarrow$ &    0.1643    & 0.1682  \\
\hline
\end{tabular}
\end{table}

\subsection{Experimental setting and main result}

\begin{figure*}[t]
    \vspace{4.75cm}
    \centering
    \parbox{0.24\linewidth}{%
    \begin{tikzpicture}[overlay,remember picture,spy using outlines={rectangle,magnification=4,size=1.6cm}]
    \node[inner sep=0pt, anchor=south west,outer sep=0pt] 
    {\pgfimage[width=\linewidth]{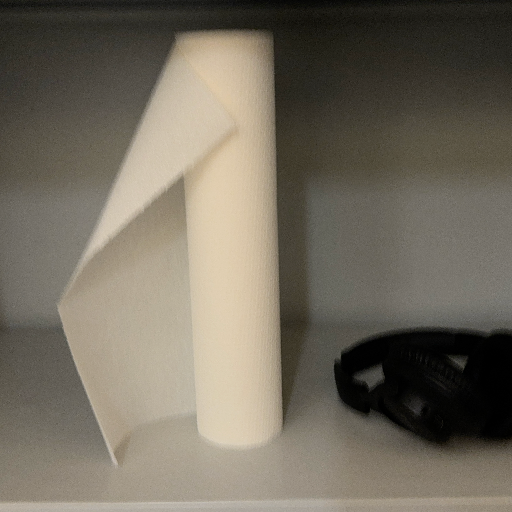}};
    {\spy[red!70!black] on (0.85,2.52) in node at (3.30,2.1);}
    \end{tikzpicture}
    }
    \parbox{0.24\linewidth}{%
    \begin{tikzpicture}[overlay,remember picture,spy using outlines={rectangle,magnification=4,size=1.6cm}]
    \node[inner sep=0pt, anchor=south west,outer sep=0pt] 
    {\pgfimage[width=\linewidth]{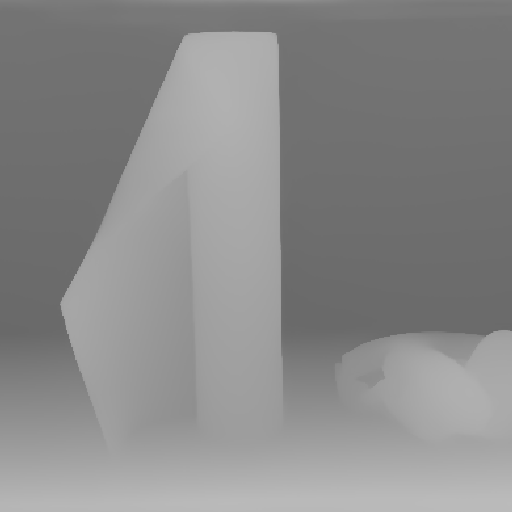}};
    {\spy[red!70!black] on (0.85,2.52) in node at (3.30,2.1);}
    \end{tikzpicture}
    }
    \parbox{0.24\linewidth}{%
    \begin{tikzpicture}[overlay,remember picture,spy using outlines={rectangle,magnification=4,size=1.6cm}]
    \node[inner sep=0pt, anchor=south west,outer sep=0pt] 
    {\pgfimage[width=\linewidth]{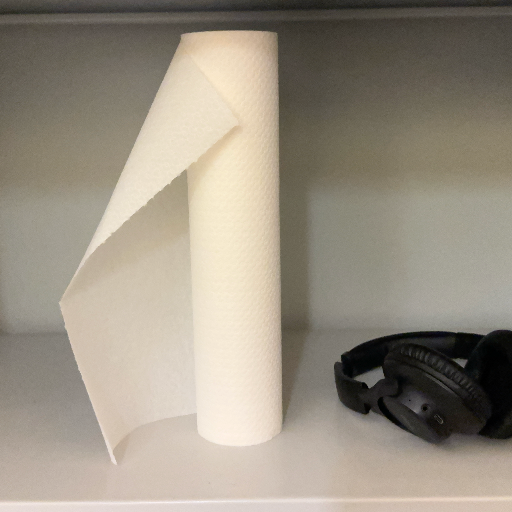}};
    {\spy[red!70!black] on (0.85,2.52) in node at (3.30,2.1);}
    \end{tikzpicture}
    }\\ \vspace{4cm}
    \parbox{0.24\linewidth}{%
    \begin{tikzpicture}[overlay,remember picture,spy using outlines={rectangle,magnification=4,size=1.6cm}]
    \node[inner sep=0pt, anchor=south west,outer sep=0pt] 
    {\pgfimage[width=\linewidth]{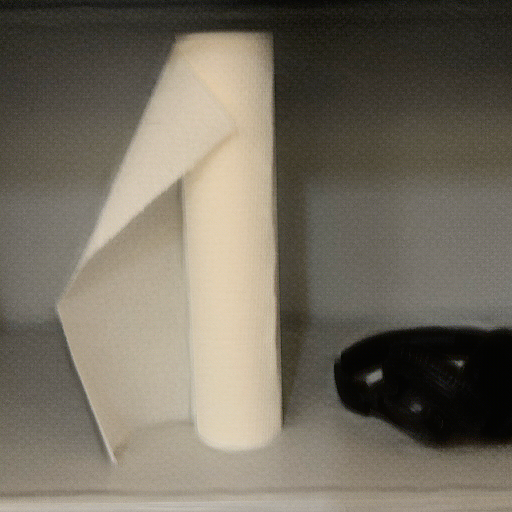}};
    {\spy[red!70!black] on (0.85,2.52) in node at (3.30,2.1);}
    \end{tikzpicture}    
    }
    \parbox{0.24\linewidth}{%
    \begin{tikzpicture}[overlay,remember picture,spy using outlines={rectangle,magnification=4,size=1.6cm}]
    \node[inner sep=0pt, anchor=south west,outer sep=0pt] 
    {\pgfimage[width=\linewidth]{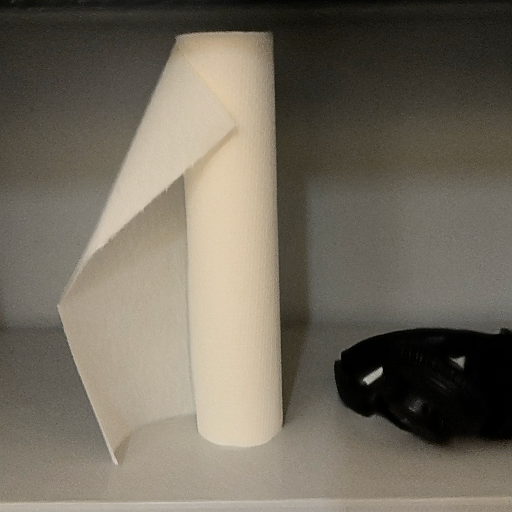}};
    {\spy[red!70!black] on (0.85,2.52) in node at (3.30,2.1);}
    \end{tikzpicture}  
    }
    \parbox{0.24\linewidth}{%
    \begin{tikzpicture}[overlay,remember picture,spy using outlines={rectangle,magnification=4,size=1.6cm}]
    \node[inner sep=0pt, anchor=south west,outer sep=0pt] 
    {\pgfimage[width=\linewidth]{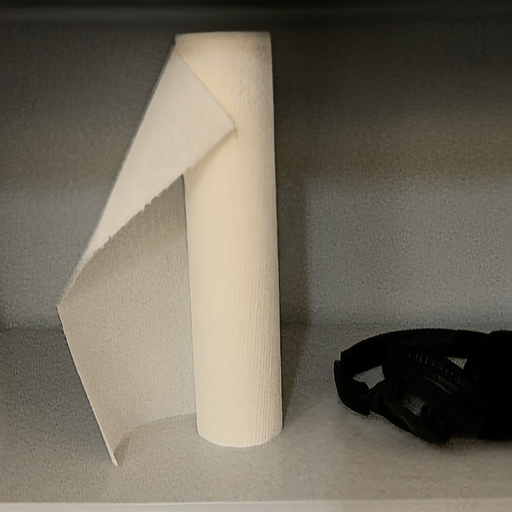}};
    {\spy[red!70!black] on (0.85,2.52) in node at (3.30,2.1);}
    \end{tikzpicture}    
    }
    \parbox{0.24\linewidth}{%
    \begin{tikzpicture}[overlay,remember picture,spy using outlines={rectangle,magnification=4,size=1.6cm}]
    \node[inner sep=0pt, anchor=south west,outer sep=0pt] 
    {\pgfimage[width=\linewidth]{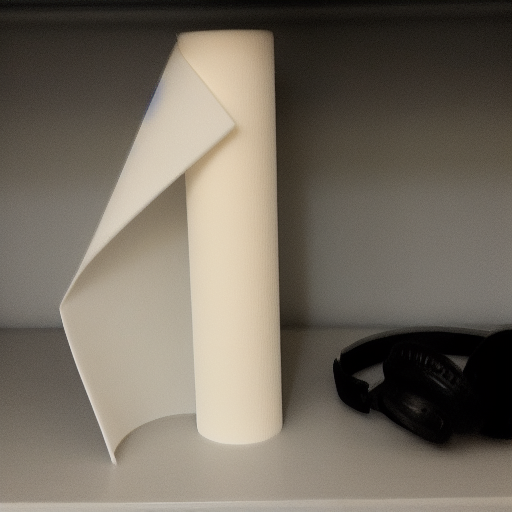}};
    {\spy[red!70!black] on (0.85,2.52) in node at (3.30,2.1);}
    \end{tikzpicture} 
    }
    \vspace{-6pt}
    \caption{Qualitative results. Top row left to right: blurred image, depth map, sharp image. Bottom row left to right: DeblurGAN, Restormer, J-MKPD, ZSLDB.}
    \label{fig:qualitative}
\end{figure*}

In this section we present deblurring results using the new dataset presented in Sec. \ref{sec:dataset}. We remark that since our focus is on real images and Lidar depth maps, this is the only existing dataset that allows such investigation. 

The proposed ZSLDB uses a pretrained Stable Diffusion with ControlNet model as image prior with depth map guidance. Notice that pretraining of SD only used generic natural images, while pretraining of ControlNet used a mixture of edges, sketches, poses, etc.. Critically, real Lidar depth maps were not needed for the pretraining process, rendering ZSLDB ``zero-shot'' in the sense that it does not require training data with real blurred images paired with depth maps and ground truths, and can be directly used on test data. We center crop and rescale the scenes in the new dataset to a standard $512 \times 512$ resolution, compatible with the SD backbone.  

We compared the proposed ZSLDB with state-of-the-art and well-known approaches to blind image deblurring. In particular, we use DeblurGAN \cite{kupyn2018deblurgan} as a classic baseline using generative models for deblurring, and Restormer \cite{zamir2022restormer} and J-MKPD \cite{carbajal2023blind} as state-of-the-art models. Quality of the deblurred images is measured by means of the LPIPS metric, which is a perceptual quality metric. We remark that a pixelwise metric like PSNR is not sufficiently robust due to imperfect alignment at a single-pixel level with respect to the ground truth. Table \ref{table:main_exp} reports the main results. We can see that ZSLDB outperforms the state-of-the-art methods by providing better image quality. Qualitative results are also shown in Fig. \ref{fig:qualitative}, where it appears that the edge information provided by the Lidar depth map allows sharper reconstruction.

\subsection{Ablation: impact of depth information}
In this section, we present an experiment to assess the impact of the depth information on the deblurring performance. In order to do this, we modify the proposed ZSLDB method by removing the ControlNet subnetwork providing depth guidance. The results are reported in Table \ref{table:depth}. It can be noticed that ZSLDB without depth guidance is competitive with the state-of-the-art methods reported in Table \ref{table:main_exp}. However, the inclusion of depth guidance results in a significant improvement in perceptual quality, highlighting its importance for image deblurring.

\subsection{Ablation: conditioning on depth edges}
Since the depth information is most useful in regularizing the deblurring problem by detecting object boundaries, we investigate whether it is more effective to provide guidance via the depth map itself or an edge map extracted from it. In particular, we use a well-known edge detection algorithm (Canny edge detector) on the depth map to derive an edge map that is provided as input to the ControlNet subnetwork. Results are reported in Table \ref{table:canny}. It can be noticed that the edge map is still effective at providing a regularizing effect but not to the same extent as using the depth map itself. We conjecture that this might mean that the depth map is providing more information than just the edges, e.g., depth gradients.

\section{Conclusions}
\label{sec:conclusions}
We presented an investigation into image deblurring guided by smartphone Lidar depth information with a focus on two contributions: the first dataset of real blurred smartphone photos with registered depth information and a novel method for zero-shot Lidar-guided deblurring based on denoising diffusion models. Results on the new dataset show that the depth information is effective at better regularizing the deblurring problem and that the proposed method outperforms state-of-the-art methods in terms of perceptual quality. 

\newpage
\ninept
\bibliographystyle{IEEEbib}

\end{document}